\documentclass[runningheads]{llncs}
\usepackage[T1]{fontenc}
\usepackage{bm}
\usepackage{marvosym}
\usepackage{multirow}
\usepackage{amsfonts,amssymb}
\usepackage{amsmath}
\usepackage{hyperref}
\hypersetup{colorlinks=true, linkcolor=blue, anchorcolor=blue, citecolor=blue, urlcolor=blue}
\usepackage{graphicx}

\begin{document}
\title{Comprehensive Generative Replay for Task-Incremental Segmentation with Concurrent Appearance and Semantic Forgetting}

\titlerunning{Comprehensive Generative Replay}
%
\newcommand*\samethanks[1][\value{footnote}]{\footnotemark[#1]}
\author{Wei Li\inst{1,3} \and
Jingyang Zhang\inst{2,4}\textsuperscript{(\Letter)}  \and
Pheng-Ann Heng\inst{2,4}  \and Lixu Gu\inst{1,3}\textsuperscript{(\Letter)} }
%

\authorrunning{W. Li et al.}
%
\institute{School of Biomedical Engineering, Shanghai Jiao Tong University, Shanghai, China \\ \email{gulixu@sjtu.edu.cn} \and
Department of Computer Science and Engineering, The Chinese University of Hong Kong, Hong Kong, China \\
\email{zjysjtu1994@gmail.com} \and
Institute of Medical Robotics, Shanghai Jiao Tong University, Shanghai, China \and Institute of Medical Intelligence and XR, The Chinese University of Hong Kong, Hong Kong, China}
\maketitle         
\begin{abstract}

Generalist segmentation models are increasingly favored for diverse tasks involving various objects from different image sources.
Task-Incremental Learning (TIL) offers a privacy-preserving training paradigm using tasks arriving sequentially, instead of gathering them due to strict data sharing policies.
However, the task evolution can span a wide scope that involves shifts in both image appearance and segmentation semantics with intricate correlation, causing concurrent appearance and semantic forgetting.
To solve this issue, we propose a Comprehensive Generative Replay (CGR) framework that restores appearance and semantic knowledge by synthesizing image-mask pairs to mimic past task data, which focuses on two aspects: modeling image-mask correspondence and promoting scalability for diverse tasks.
Specifically, we introduce a novel Bayesian Joint Diffusion (BJD) model for high-quality synthesis of image-mask pairs with their correspondence explicitly preserved by conditional denoising.
Furthermore, we develop a Task-Oriented Adapter (TOA) that recalibrates prompt embeddings to modulate the diffusion model, making the data synthesis compatible with different tasks.
Experiments on incremental tasks (cardiac, fundus and prostate segmentation) show its clear advantage for alleviating concurrent appearance and semantic forgetting.
Code is available at \url{https://github.com/jingyzhang/CGR}.

\end{abstract}
\section{Introduction}
For medical image segmentation, there is an increasing demand for a generalist model capable of handling diverse tasks, involving multiple objectives in various imaging conditions \cite{ma2024segment,huang2023stu,liu2023clip}.
As data for different tasks is often dispersed across clinical departments with stringent data sharing policies \cite{price2019privacy}, aggregating these task data as a consolidated set is impractical for model training \cite{liu2023incremental} while Task-Incremental Learning (TIL) emerges as a storage-efficient and privacy-protecting paradigm.
It enables a generalist model to learn from sequentially arriving tasks, where, notably, \emph{these tasks vary widely without predefined restrictions}, addressing diverse objectives even across different anatomical regions and imaging devices.

A straightforward TIL approach is to consecutively finetune the model using only the current task, yet leading to dramatic failure on previously learned tasks, attributed to:
1) co-occurrence of appearance discrepancy and semantic shift, as tasks can evolve in a wide scope that involves not only heterogeneous data from multiple resources but also varying objectives with distinct anatomy \cite{ma2024segment};
and 2) ignorance of appearance-semantics correspondence, making the model's memorizability even fragile to such drastic task shifts \cite{li2022domain}.
We identify this phenomenon as \emph{concurrent appearance and semantic forgetting}, which is under-studied in TIL.

However, in contrast to this challenging TIL scenario where appearance and semantic forgetting are intrinsically coupled, prevalent research recognizes these forgetting problems separately, addressing each by customized paradigms.
Specifically, Class-Incremental Learning (CIL) alleviates semantic forgetting for an increasing variety of segmentation objectives within a confined region \cite{liu2023incremental} through distilling \cite{douillard2021plop,zhao2023inherit} and transferring \cite{liu2022learning,wu2023continual} semantic prototypes.
Nevertheless, in the TIL context, where objectives often span entirely different regions with substantial appearance discrepancies, CIL methods would suffer intractable appearance forgetting that hinders the reliability of semantic transfer and distillation \cite{liu2023incremental,douillard2021plop}.
In addition, Domain-Incremental Learning (DIL) combats appearance forgetting for a deterministic objective via style regularization \cite{li2017learning,kirkpatrick2017overcoming,zhang2023s} and appearance recovery using image-only generative replay with error-prone mask reuse \cite{li2022domain,chen2023generative}.
Yet, in TIL with varying objectives, such intricate semantic shift disrupts the style-oriented regularization and even makes the image-only replay dominated by the current task objective \cite{li2022domain}, leading to semantic forgetting in DIL methods for past objectives.
Overall, while existing DIL and CIL effectively tackle either appearance or semantic forgetting individually, they fall short of a comprehensive TIL perspective that simultaneously considers both issues with high correspondence.

Based on the above issues, compared with traditional DIL and CIL schemes, the main challenge in TIL lies in how to construct a unified framework that comprehensively overcomes concurrent appearance and semantic forgetting, rather than treating them separately.
Inspired by the practical data rehearsal \cite{zhang2022learning}, our insight involves synthesizing both images and their corresponding segmentation masks to simulate diverse past task data, which serves as a comprehensive replay mechanism to recover the coupled image appearance and segmentation semantics for memory evoking.
Recently, diffusion models \cite{ho2020denoising} have provided a cutting-edge image synthesis approach especially for the medical field \cite{muller2022diffusion}, yet they are limited in simultaneously synthesizing the corresponding segmentation masks due to the ignorance of essential image-mask correlation \cite{bao2023one}.
Moreover, diffusion models exhibit a decent versatility in switching between synthesis tasks under the guidance of text prompts \cite{rombach2022high} with Contrastive Language-Image Pretraining (CLIP)-based embedding \cite{radford2021learning}.
However, this pretrained embedding poses a significant domain gap for our customized medical context \cite{liu2023clip}, misguiding the diffusion model and impeding its scalable data synthesis for replaying diverse tasks.
These insights motivate us to pursue comprehensive replay using a diffusion model, focusing on two aspects:
preserving crucial image-mask correspondence, and regulating the CLIP-based embedding to make data synthesis compatible with diverse tasks.

In this paper, we present \emph{to our knowledge the first TIL paradigm} for medical image segmentation, accommodating a wide task scope with diverse objectives.
Our contributions are three-fold:
1) We propose a Comprehensive Generative Replay (CGR) framework to reduce concurrent appearance and semantic forgetting across diverse tasks, by generating image-mask pairs to reproduce past task data;
2) We design a novel Bayesian Joint Diffusion (BJD) model for structure-realistic synthesis of image-mask pairs, formulating their correspondence as conditional distributions and optimizing through conditional denoising;
and 3) We propose a Task-Oriented Adapter (TOA) that recalibrates the CLIP-based embedding to modulate the diffusion model, promoting synthesis scalability for diverse tasks.
We evaluate our method on tasks for cardiac, fundus, and prostate segmentation, showing its minimal forgetting and clear advantages over DIL and CIL methods.

\begin{figure}[t]
    \includegraphics[width=\textwidth]{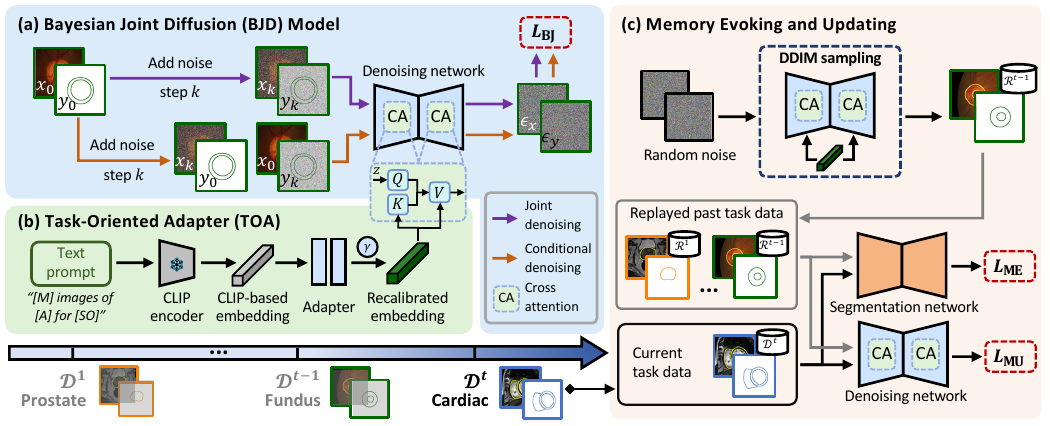}
    \caption{Illustration of our proposed Comprehensive Generative Replay (CGR) framework for task-incremental learning, e.g., on prostate, fundus, and cardiac segmentation.
    Specifically, we synthesize paired images and segmentation masks to simulate past task data,
    by adopting a Bayesian Joint Diffusion (BJD) model to preserve image-mask correspondence (Sec. \ref{sec:BJD}),
    and equipping a Task-Oriented Adapter (TOA) on the CLIP-based embedding to modulate the diffusion model for scalable data synthesis (Sec. \ref{sec:TOA}).
    When encountering a new task, we leverage replayed past task data to evoke the faded memory, and update it to include this new task knowledge for future replays (Sec. \ref{sec:replay}).
    }
    \label{fig:overall}
\end{figure}

\section{Method}
In the Task-Incremental Learning (TIL) scenario, we train a segmentation model $f_{\mathrm{\theta}}$ with sequentially arriving tasks.
At each learning round $t$, we only obtain data from the current task $\mathcal{D}^t=\{\mathcal{X}^t,\mathcal{Y}^t\}$ of paired images and segmentation masks, without access to past tasks $\{\mathcal{D}^i\}_{i=1}^{t-1}$.
Fig.~\ref{fig:overall} illustrates our Comprehensive Generative Replay (CGR) framework to reduce concurrent appearance and semantic forgetting, as $\mathcal{X}^t$ and $\mathcal{Y}^t$ both have diverse distributions under drastic task shifts.

\subsection{Bayesian Joint Diffusion Model}
\label{sec:BJD}
To combat concurrent appearance and semantic forgetting in TIL, we synthesize paired images and segmentation masks to mimic past task inputs using diffusion models \cite{ho2020denoising}.
To achieve this, a naive way is to directly model the joint distribution of images and masks \cite{zhang2023jointnet}, but their correspondence is often disrupted and even ignored \cite{bao2023one}.
Therefore, we propose a Bayesian framework to preserve image-mask correspondence, ensuring precise semantics well-aligned in the recovered images.

\noindent
\textbf{Naive Joint Diffusion (NJD) Model.}
To synthesize paired images and masks using a diffusion model, a basic idea is to learn their joint distribution through applying forward and reverse processes \cite{ho2020denoising} on image-mask pairs \cite{bao2023one}.
It should be performed in each previous learning round to simulate past task data $\mathcal{D}^{i\in[1:t-1]}$.
Specifically, in the forward process, given an original image-mask pair $(x_0,y_0)$ of the task $\mathcal{D}^{i}=\{\mathcal{X}^i,\mathcal{Y}^i\}$, we gradually add Gaussian noise to it, and achieve noisy pairs $\{(x_k,y_k)\}_{k=1}^K$ over $K$ steps for training a denoising network $\epsilon_\theta$.
Then, this $\epsilon_\theta$ is utilized in the reverse process to iteratively recover $(x_0,y_0)$ from $(x_K,y_K)$.

The denoising network $\epsilon_\theta$ can be trained with the Maximum Likelihood Estimation (MLE) on the image-mask joint distribution, i.e., $\mathop{\mathrm{max}} \log p ({x},{y})$.
It could be simplified for the joint denoising score matching \cite{bao2023one}, under a naive assumption that images and masks are nearly independent \cite{zhang2023jointnet} with their correlation being corrupted by noise added in the forward process.
Specifically, in the forward step $k$, noise ${\epsilon}_{x}$ and ${\epsilon}_{y}$ are added to $x_0$ and $y_0$ of the task $\mathcal{D}^i$, respectively, leading to ${x}_k = \sqrt{\bar{\alpha}_k} {x}_0 + \sqrt{1-\bar{\alpha}_k} {\epsilon}_x$ and ${y}_k = \sqrt{\bar{\alpha}_k} {y}_0 + \sqrt{1-\bar{\alpha}_k} {\epsilon}_y$ with a noise level $\bar{\alpha}_k$.
Then, given these as inputs, $\epsilon_\theta$ is trained to jointly predict both noise as a multi-channel output $[{\epsilon}_{x},{\epsilon}_{y}]$ to recover the image-mask distribution in this task:
\begin{equation}
    L_{\text{NJ}}^{\mathcal{D}^i}(\epsilon_\theta) = {\mathbb{E}}_{k,\epsilon_x,\epsilon_y,(x_0,y_0)\sim\mathcal{D}^i}\Big [ \Vert [{\epsilon}_{x},{\epsilon}_{y}] - {\epsilon}_\theta ({x}_k, {y}_k, k) \Vert^2_2 \Big ].
    \label{eq:L_NJD}
\end{equation}
However, the correspondence between images and masks is easily distorted by the random noise added simultaneously to both elements, especially with noise levels increasing gradually in the forward process.
It can cause misaligned appearance and semantics, hampering the synthesis of structure-realistic image-mask pairs.

\noindent
\textbf{Bayesian Joint Diffusion (BJD) Model.}
To solve this problem, we propose a Bayesian framework that leverages conditional distributions to model the image-mask correspondence.
To this end, the MLE objective can be reformulated as:
\begin{equation}
\label{eq:bj-objective}
    \mathop{\mathrm{max}} \log p (x,y) =
     \mathop{\mathrm{max}}
     \Big [\log p(x)p(y) + \log p(x|y) + \log p(y|x) \Big ],
\end{equation}
where $\mathop{\mathrm{max}}\log p(x)p(y)$ can be regarded as a degenerated NJD case \cite{zhang2023jointnet} under the assumption of independence with correspondence being distorted, converted to a similar joint denoising as Eq.~(\ref{eq:L_NJD}).
Moreover, $\mathop{\mathrm{max}}\log p(x|y)$ and $\mathop{\mathrm{max}}\log p(y|x)$ for conditional distributions capture interactions across images and masks, enforcing alignment between appearance and semantics.
Notably, they could be simplified for conditional denoising \cite{ho2022classifier}, where the noise-free mask $y_0$ and image $x_0$ are used alternatively as reliable references for clear unidirectional correspondence, rather than adding noise simultaneously that would corrupt directional correspondence:
\begin{align}
    L_{\text{BJ}}^{\mathcal{D}^i}(\epsilon_\theta) = {\mathbb{E}}_{k,\epsilon_x,\epsilon_y,(x_0,y_0)\sim\mathcal{D}^i}\Big [ & \Vert [{\epsilon}_{x},{\epsilon}_{y}] - {\epsilon}_\theta ({x}_k, {y}_k, k) \Vert^2_2 +
    \Vert [{\epsilon}_{x},{0}] - {\epsilon}_\theta ({x}_k, {y}_0, k) \Vert^2_2 \nonumber\\
    &+\Vert [{0},{\epsilon}_{y}] - {\epsilon}_\theta ({x}_0, {y}_k, k) \Vert^2_2 \Big ].
    \label{eq:L_BJD}
\end{align}
Intuitively, conditional image denoising with $\epsilon_\theta(x_k,y_0,k)$ restores image appearance that aligns with semantics from the clear mask $y_0$, while conditional mask denoising with $\epsilon_\theta(x_0,y_k,k)$ captures accurate semantics in the clean image $x_0$. Although conditional denoising incurs higher FLOPs during training, it does not involve extra learnable parameters, avoiding higher FLOPs during inference.

\subsection{Task-Oriented Adapter (TOA)}
\label{sec:TOA}
In our TIL scenario, BJD needs to scale across diverse data distributions of previous tasks, in order to simulate each task effectively.
To ensure such versatility, a practical approach is to modulate the diffusion model with text prompts encoded as CLIP-based embedding \cite{rombach2022high}.
However, this embedding is pretrained on natural language-image databases \cite{radford2021learning} and may not be compatible with our customized tasks \cite{liu2023clip}.
To address this problem, we propose a Task-Oriented Adapter (TOA) to recalibrate the prompt embedding, and leverage a cross attention mechanism to modulate the BJD model for scalable data synthesis across diverse tasks.

\noindent
\textbf{TOA Architecture.}
We establish a text prompt template \cite{liu2023clip} for task characterization: ``[M] images of [A] for [SO]'', where [M], [A] and [SO] represent the modality, region, and objective, respectively.
Given a task with index $i$, the text prompt is fed into the CLIP encoder \cite{radford2021learning} to obtain the prompt embedding $e^i$, which is further recalibrated by a task-specific, lightweight two-layer adapter $\psi_\theta^i$, ensuring scalability for various tasks in a memory-efficient way:
\begin{equation}
    w^i = e^i + \gamma^i \psi_\theta^i(e^i),
\end{equation}
where $\gamma^i$ is a learnable factor to modify the recalibration strength.

\noindent
\textbf{Modulated Denoising Network.}
The recalibrated embedding $w^i$ is utilized to modulate BJD for task-specific data synthesis.
We combine $w^i$ with the output $z_j$ from each intermediate layer $j$ of the denoising network $\epsilon_\theta$ using cross-attention $\tau_j(\cdot)$ \cite{vaswani2017attention}, assembling them into a modulated denoising network $\mu_\theta$ for BJD:
\begin{equation}
\mu_\theta = \{\tau_j(w^i,z_j)\}_j,  \ \ \text{where} \ \  \tau_j(w^i,z_j)=\sigma(\dfrac{(Q_jz_j)\cdot(K_jw^i)^T}{\sqrt{d_j}})\cdot(V_jw^i).
\end{equation}
Here, $Q_j$, $K_j$, and $V_j$ are learnable query, key, and value for cross attention in each layer $j$.
Moreover, $d_j$ is the channel number and $\sigma(\cdot)$ is the softmax function.

\subsection{Replay for Memory Evoking and Updating}
\label{sec:replay}
Based on the BJD with TOA, we perform DDIM sampling \cite{song2020denoising} to generate image-mask pairs for replaying past task data, denoted as $\mathcal{R}^{i\in[1:t-1]}$.
The replayed data evokes faded memory when learning on a new task, and this memory should also be updated to include the new task for subsequent replay in future rounds \cite{chen2023generative}.

\noindent
\textbf{Memory Evoking.}
To evoke previous memory when learning on a new task, we take the replayed past task datasets $\mathcal{R}^{i\in[1:t-1]}$ to jointly train the segmentation model $f_\theta$ together with the current task data $\mathcal{D}^t$, where the segmentation loss $L_{\text{seg}}$ is calculated for $\mathcal{D}^t$ and $\mathcal{R}^i$ using cross-entropy with a trade-off $\alpha$:
\begin{equation}
L_{\text{ME}}(f_\theta) =  L^{\mathcal{D}^t}_{\text{seg}}(f_\theta) + \alpha{\sum}_{i=1}^{t-1} L^{\mathcal{R}^i}_{\text{seg}}(f_\theta).
\end{equation}

\noindent
\textbf{Memory Updating.}
In addition to mimicking past tasks, the memory-evoking diffusion model should also simulate the current task data for subsequent replays in TIL.
Therefore, we update the BJD model using loss functions measured on $\mathcal{D}^t$ and $\mathcal{R}^{i}$ with a trade-off $\beta$, while equipping the modulated denoising network $\mu_\theta$ through TOA for scalable data synthesis:
\begin{equation}
L_{\text{MU}}(\mu_\theta) = L_{\text{BJ}}^{\mathcal{D}^t}(\mu_\theta) + \beta{\sum}_{i=1}^{t-1} L_{\text{BJ}}^{\mathcal{R}^i}(\mu_\theta).
\end{equation}

\section{Experiments}

\noindent
\textbf{Dataset.}
We evaluated our method on three tasks:
1) cardiac MRI segmentation \cite{campello2021multi} with 320 subjects for three structures of the left ventricle, the right ventricle, and the left ventricular myocardium;
2) fundus segmentation \cite{wang2020dofe} with 1060 subjects for the optic cup and disc;
and 3) prostate MRI segmentation \cite{liu2020ms} with 116 subjects.
They present concurrent appearance and semantic shifts due to varying imaging conditions and diverse objectives.
For data pre-processing, cardiac and prostate images were resampled to unit spacing and resized to 256$\times$256 in the axis plane, while fundus images were cropped to 800$\times$800 around the optic disc and resized to the same size 256$\times$256. In each task, the dataset was split into 60\%, 15\%, and 25\% for training, validation, and testing, respectively.

\noindent
\textbf{Experimental Setting.}
We organized these segmentation tasks in two learning schedules: $S_{C\to F\to P}$ with data arriving sequentially for cardiac, fundus and prostate segmentation, and $S_{P\to F\to C}$ as a reverse order.
After completing learning on the final task, we evaluated the model on current and previous tasks using the Dice Score Coefficient (DSC) and the $95\%$ Hausdorff Distance (HD).

\noindent
\textbf{Implementation.}
We used 2D Res-UNet as our segmentation backbone due to its strong generalist transferability \cite{huang2023stu}.
It was trained using Adam Optimizer with learning rate 2$\times$10$^{-4}$, batch size 16, and iteration number 4$\times$10$^{4}$.
For BJD with $K=1000$ forward steps, we employed a 2D-UNet as the denoising network \cite{ho2020denoising}, using AdamW Optimizer with learning rate 1$\times$10$^{-4}$, batch size 8 and iteration number 4$\times$10$^{5}$. We empirically synthesized 3000 image-mask pairs for each previous task.
We empirically set $\alpha$ and $\beta$ as 0.25 for a suitable trade-off \cite{li2022domain}.

\begin{table}[t]
    \caption{Performance comparison after learning consecutively on Cardiac (C), Fundus (F), and Prostate (P) segmentation tasks, with two different learning orders $S_{C\to F \to P}$ and $S_{P\to F \to C}$. \textbf{Bold} font denotes the best performance in this learning process (except JointTrain which performs offline and serves as the upper bound).}
    \resizebox{\linewidth}{!}{
    \begin{tabular}{ll|cccccccc|cccccccc}
        \hline \hline
        \multicolumn{2}{l|}{Learning order} & \multicolumn{8}{c|}{$S_{C\to F \to P}$ (Cardiac $\to$ Fundus $\to$ Prostate)} & \multicolumn{8}{c}{$S_{P\to F \to C}$ (Prostate $\to$ Fundus $\to$ Cardiac)} \\
        \hline \hline

        \multicolumn{2}{l|}{\multirow{2}{*}{Task}} & \multicolumn{2}{c|}{Previous} & \multicolumn{1}{c|}{Current} & \multicolumn{1}{c|}{\multirow{2}{*}{Mean}} & \multicolumn{2}{c|}{Previous} & \multicolumn{1}{c|}{Current} & \multirow{2}{*}{Mean} & \multicolumn{2}{c|}{Previous} & \multicolumn{1}{c|}{Current} & \multicolumn{1}{c|}{\multirow{2}{*}{Mean}} & \multicolumn{2}{c|}{Previous} & \multicolumn{1}{c|}{Current} & \multirow{2}{*}{Mean} \\ \cline{3-5} \cline{7-9} \cline{11-13} \cline{15-17}

        \multicolumn{2}{l|}{} & Cardiac & \multicolumn{1}{c|}{Fundus} & \multicolumn{1}{c|}{Prostate} & \multicolumn{1}{c|}{} & Cardiac & \multicolumn{1}{c|}{Fundus} & \multicolumn{1}{c|}{Prostate} &  & Prostate & \multicolumn{1}{c|}{Fundus} & \multicolumn{1}{c|}{Cardiac} & \multicolumn{1}{c|}{} & Prostate & \multicolumn{1}{c|}{Fundus} & \multicolumn{1}{c|}{Cardiac} &  \\
        \hline \hline

        \multicolumn{2}{l|}{Metric} & \multicolumn{4}{c|}{\textbf{DSC (\%) $\uparrow$}} & \multicolumn{4}{c|}{\textbf{HD (pixel) $\downarrow$}} & \multicolumn{4}{c|}{\textbf{DSC (\%) $\uparrow$}} & \multicolumn{4}{c}{\textbf{HD (pixel) $\downarrow$}} \\
        \hline \hline

        \multicolumn{1}{l|}{\multirow{2}{*}{BaseLine }} & { JointTrain} & 89.79 & \multicolumn{1}{c|}{89.34} & \multicolumn{1}{c|}{90.81} & \multicolumn{1}{c|}{89.98} & 2.02 & \multicolumn{1}{c|}{5.10} & \multicolumn{1}{c|}{3.36} & 3.49 & 90.81 & \multicolumn{1}{c|}{89.34} & \multicolumn{1}{c|}{89.79} & \multicolumn{1}{c|}{89.98} & 3.36 & \multicolumn{1}{c|}{5.10} & \multicolumn{1}{c|}{2.02} & 3.49 \\

        \multicolumn{1}{l|}{} & { FineTune} & 0.03 & \multicolumn{1}{c|}{0.02} & \multicolumn{1}{c|}{\textbf{91.06}} & \multicolumn{1}{c|}{30.37} & NaN & \multicolumn{1}{c|}{NaN} & \multicolumn{1}{c|}{\textbf{2.69}} & \multicolumn{1}{c|}{NaN} & 0.00 & \multicolumn{1}{c|}{0.00} & \multicolumn{1}{c|}{\textbf{89.60}} & \multicolumn{1}{c|}{29.87} & NaN & \multicolumn{1}{c|}{NaN} & \multicolumn{1}{c|}{2.11} & {NaN} \\
        \hline

        \multicolumn{1}{l|}{\multirow{3}{*}{DIL}} & { EWC \cite{kirkpatrick2017overcoming}}  &  21.84 & \multicolumn{1}{c|}{49.41} & \multicolumn{1}{c|}{73.37} & \multicolumn{1}{c|}{48.20} & 139.89 & \multicolumn{1}{c|}{25.88} & \multicolumn{1}{c|}{15.31} & \multicolumn{1}{c|}{60.36} & 0.27 & \multicolumn{1}{c|}{44.59} & \multicolumn{1}{c|}{58.47} & \multicolumn{1}{c|}{34.44} & 151.96 & \multicolumn{1}{c|}{62.38} & \multicolumn{1}{c|}{38.84} & 84.40 \\

        \multicolumn{1}{l|}{} & { LwF \cite{li2017learning}} & 46.76 & \multicolumn{1}{c|}{82.37} & \multicolumn{1}{c|}{88.08} & \multicolumn{1}{c|}{72.40} & 12.95 & \multicolumn{1}{c|}{9.44} & \multicolumn{1}{c|}{6.49} & \multicolumn{1}{c|}{9.63} & 19.34 & \multicolumn{1}{c|}{77.89} & \multicolumn{1}{c|}{86.94} & \multicolumn{1}{c|}{61.39} & 11.95 & \multicolumn{1}{c|}{12.50} & \multicolumn{1}{c|}{3.89} & 9.45 \\

        \multicolumn{1}{l|}{} & { GAR \cite{chen2023generative}} & 86.12 & \multicolumn{1}{c|}{84.39} & \multicolumn{1}{c|}{90.30} & \multicolumn{1}{c|}{86.94} & 4.63 & \multicolumn{1}{c|}{11.76} & \multicolumn{1}{c|}{2.93} & \multicolumn{1}{c|}{6.44} & 73.27 & \multicolumn{1}{c|}{84.51} & \multicolumn{1}{c|}{89.19} & \multicolumn{1}{c|}{82.33} & 5.61 & \multicolumn{1}{c|}{8.38} & \multicolumn{1}{c|}{2.18} & 5.39 \\
        \hline

        \multicolumn{1}{l|}{\multirow{3}{*}{CIL}} & { PLOP \cite{douillard2021plop}} & 22.47 & \multicolumn{1}{c|}{72.39} & \multicolumn{1}{c|}{81.96} & \multicolumn{1}{c|}{58.94} & 32.54 & \multicolumn{1}{c|}{15.93} & \multicolumn{1}{c|}{13.21} & \multicolumn{1}{c|}{20.56} & 68.62 & \multicolumn{1}{c|}{79.16} & \multicolumn{1}{c|}{85.18} & \multicolumn{1}{c|}{77.65} & 5.25 & \multicolumn{1}{c|}{13.53} & \multicolumn{1}{c|}{5.10} & 7.96 \\

        \multicolumn{1}{l|}{} & { MEIL \cite{liu2022learning}}  & 39.84 & \multicolumn{1}{c|}{74.05} & \multicolumn{1}{c|}{88.20} & \multicolumn{1}{c|}{67.36} & 20.68 & \multicolumn{1}{c|}{12.30} & \multicolumn{1}{c|}{5.84}& \multicolumn{1}{c|}{12.94} & 4.51 & \multicolumn{1}{c|}{41.42} & \multicolumn{1}{c|}{86.54} & \multicolumn{1}{c|}{44.16} & 24.34 & \multicolumn{1}{c|}{73.17} & \multicolumn{1}{c|}{3.98} & 33.83 \\

        \multicolumn{1}{l|}{} & { HSI \cite{liu2023incremental}}  & 85.40 & \multicolumn{1}{c|}{81.96} & \multicolumn{1}{c|}{89.78} & \multicolumn{1}{c|}{85.72} & 5.56 & \multicolumn{1}{c|}{10.82} & \multicolumn{1}{c|}{2.77} & \multicolumn{1}{c|}{6.38} & 67.60 & \multicolumn{1}{c|}{81.27} & \multicolumn{1}{c|}{89.24} & \multicolumn{1}{c|}{79.37} & 7.89 & \multicolumn{1}{c|}{14.70} & \multicolumn{1}{c|}{2.09} & 8.23  \\
        \hline

        \multicolumn{1}{l|}{\multirow{3}{*}{TIL}} & { CGR (-BJD) } & 85.43 & \multicolumn{1}{c|}{86.02} & \multicolumn{1}{c|}{90.28} & \multicolumn{1}{c|}{87.24} & 3.89 & \multicolumn{1}{c|}{7.16} & \multicolumn{1}{c|}{2.74} & \multicolumn{1}{c|}{4.60} & 83.02 & \multicolumn{1}{c|}{85.36} & \multicolumn{1}{c|}{89.34} & \multicolumn{1}{c|}{85.91} & 5.52 & \multicolumn{1}{c|}{7.22} & \multicolumn{1}{c|}{2.02} & 4.92\\

        \multicolumn{1}{l|}{} & { CGR (-TOA) } & 86.77 & \multicolumn{1}{c|}{86.78} & \multicolumn{1}{c|}{90.47} & \multicolumn{1}{c|}{88.01} & 3.50 & \multicolumn{1}{c|}{6.62} & \multicolumn{1}{c|}{3.26} & \multicolumn{1}{c|}{4.46} & 85.08 & \multicolumn{1}{c|}{85.89} & \multicolumn{1}{c|}{89.31} & \multicolumn{1}{c|}{86.76} & 4.72 & \multicolumn{1}{c|}{7.16} & \multicolumn{1}{c|}{\textbf{2.00}} & 4.63 \\

        \multicolumn{1}{l|}{} & { CGR (ours)} & \textbf{87.48} & \multicolumn{1}{c|}{\textbf{87.99}} & \multicolumn{1}{c|}{90.68} & \multicolumn{1}{c|}{\textbf{88.71}} & \textbf{2.81} & \multicolumn{1}{c|}{\textbf{5.83}} & \multicolumn{1}{c|}{2.93} & \multicolumn{1}{c|}{\textbf{3.86}} & {\textbf{88.00}} & \multicolumn{1}{c|}{\textbf{87.40}} & \multicolumn{1}{c|}{89.31} & \multicolumn{1}{c|}{\textbf{88.24}} & \textbf{4.68} & \multicolumn{1}{c|}{\textbf{6.06}} & \multicolumn{1}{c|}{2.07} & \textbf{4.27} \\
        \hline \hline
    \end{tabular}
    }
    \label{tab:comparison}
\end{table}

\begin{figure}[t]
    \includegraphics[width=\textwidth]{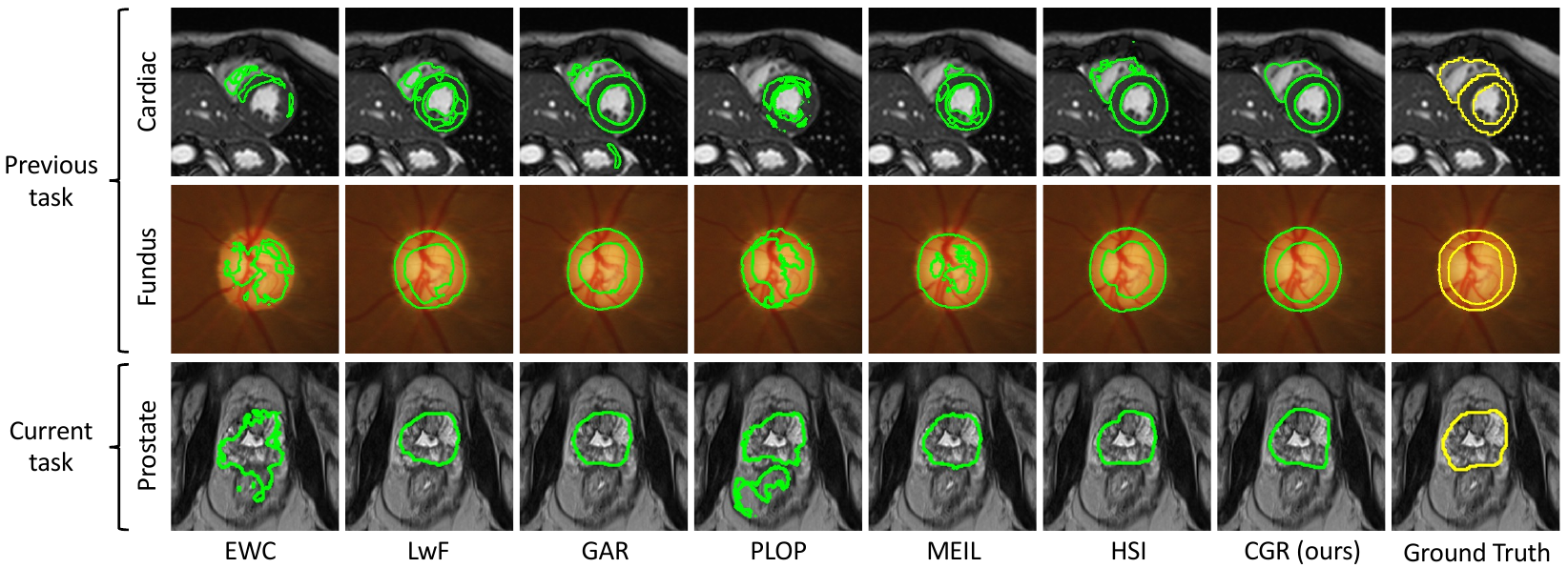}
    \caption{Segmentation examples by learning on sequentially arriving tasks for cardiac, fundus, and prostate segmentation.}
    \label{fig:seg}
\end{figure}

\noindent
\textbf{Comparison with State-of-the-Arts.}
We compared our TIL framework called Comprehensive Generative Replay (\textbf{CGR}), including a Bayesian Joint Diffusion (\textbf{BJD}) model and a Task-Oriented Adapter (\textbf{TOA}), with state-of-the-art methods:
1) Baselines: \textbf{JointTrain} that gathers all task data for joint training, and \textbf{FineTune} that sequentially finetunes the model using only current task data;
2) DIL schemes: \textbf{EWC} \cite{kirkpatrick2017overcoming} and \textbf{LwF} \cite{li2017learning} using appearance regularization terms, and \textbf{GAR} \cite{chen2023generative} based on generative appearance replay without concurrent semantics synthesis;
and 3) CIL schemes: \textbf{PLOP} \cite{douillard2021plop} using class distillation and \textbf{MEIL} \cite{liu2022learning} transferring class prototypes,
and \textbf{HSI} \cite{liu2023incremental} alleviating semantic shift under slight data heterogeneity, yet halting in an extremely simple TIL case restricted in an enclosed region, unlike our scenario with a broader task range.

As presented in Table~\ref{tab:comparison}, FineTune fails drastically for previous tasks, yielding 0.00\% DSC and NaN HD due to concurrent appearance and semantic forgetting during task evolution.
Although DIL and CIL schemes can outperform FineTune, their advantages are modest, addressing forgetting problems partially for either appearance or semantics.
Notably, GAR achieves a relatively higher performance in the DIL scheme by predicting segmentation masks on the replayed images to generate restored yet error-prone semantics \cite{li2022domain}, while HSI stands out in the CIL scheme owing to its dual-flow module with batch renormalization layers to resist to mild image heterogeneity alongside semantic shift.
However, their performance is still hindered in the TIL scenario, where tasks arrive with not only significant data discrepancy but also entirely different objectives.
Our CGR shows the best performance closer to JointTrain, e.g. substantially outperforming GAR and HSI with 1.77\% and 2.99\% DSC for the learning order $S_{C\to F\to P}$, demonstrating its clear advantages in addressing both appearance and semantic forgetting for TIL.

Furthermore, Fig.~\ref{fig:seg} visualizes the segmentation results for previous and current tasks.
Most methods exhibit sufficient adaptation capability on the current task, except EWC and PLOP with over-strong regularization terms.
Our CGR achieves the precise shape and smooth boundary, especially for previous tasks, whereas other methods show considerable limitations.

\begin{figure}[t]
    \includegraphics[width=\textwidth]{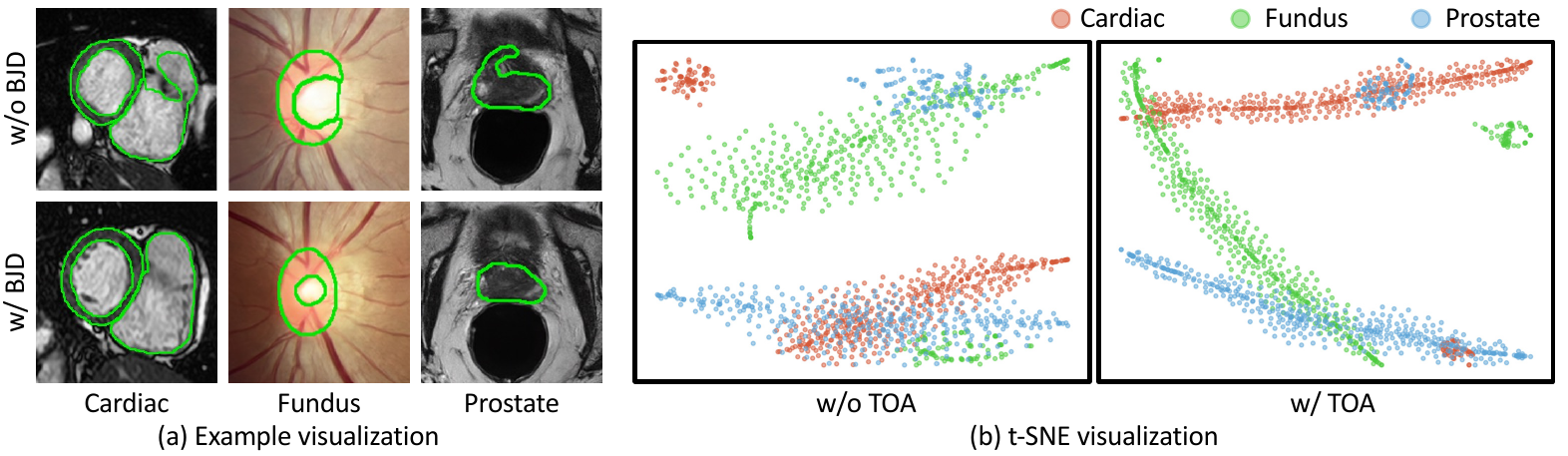}
    \caption{Ablation analysis. Synthesized image-mask pairs are displayed in (a) with and without BJD, and their t-SNE visualization is shown in (b) with and without TOA.}
    \label{fig:ablation}
\end{figure}

\noindent
\textbf{Effectiveness of Bayesian Joint Diffusion (BJD).}
We visualize in Fig.~\ref{fig:ablation}(a) the synthesized image-mask pairs for cardiac, fundus and prostate segmentation.
The image-mask pairs generated without BJD, i.e., using the Naive Joint Diffusion (NJD) instead, often suffer irregular structure fracture and disruption.
BJD provides the structure-realistic data synthesis with normal shapes preserved for different segmentation objectives, significantly improving our CGR method over the BJD-deactivated counterpart (-BJD), as shown in Table~\ref{tab:comparison} for ablation study.

\noindent
\textbf{Contribution of Task-Oriented Adapter (TOA).}
In Fig.~\ref{fig:ablation}(b), we use t-SNE to visualize the distributions of synthesized image-mask pairs with and without TOA attached to the BJD model.
TOA enhances the compactness of inner-task distribution and the separability of inter-task distribution, promoting scalability of data synthesis across tasks.
Integrating TOA in CGR has superiority over the TOA-deactivated counterpart (-TOA), as shown in Table~\ref{tab:comparison} for ablation study.

\noindent
\textbf{Robustness to Learning Order.}
We validate the robustness of our method to variations in the learning order, as shown in Table~\ref{tab:comparison} using forward and reversed orders.
Other methods exhibit unstable performance and significant fluctuations under different learning orders.
Notably, our CGR consistently outperforms these methods by a stable and substantial margin, highlighting its strong robustness.

\section{Conclusion}
This paper presents a novel TIL framework for medical image segmentation, addressing concurrent appearance and semantic forgetting as tasks evolve widely.
To achieve this, we propose a Comprehensive Generative Replay (CGR) framework that synthesizes past task data including paired images and masks, where their correspondence is captured by a Bayesian Joint Diffusion (BJD) model with Task-Oriented Adapter (TOA) for synthesis scalability.
The broader significance of our work may lie in the extension on foundation models, providing a promising avenue for accumulating generalist skills in artificial general intelligence.
In the future, we will conduct more comprehensive experiments to validate the robustness of CGR on multi-organ or tumor segmentation datasets.

\begin{credits}
\subsubsection{\ackname} 
This research is partially supported by Specific Project of Shanghai Jiao Tong University for "Invigorating Inner Mongolia through Science and Technology" (2022XYJG0001–01–17), the funding from Star of SJTU Programme, a grant from the Researh Grants Council of the
Hong Kong Special Administrative Region, China (Project No.: T45-401/22-N), and a grant
from Hong Kong Innovation and Technology Fund (Project No.: MHP/085/21).

\subsubsection{\discintname}
The authors have no competing interests to declare that are
relevant to the content of this article.

\end{credits}

\bibliographystyle{splncs04}
\bibliography{Paper-0187}

\begin{thebibliography}{10}
\providecommand{\url}[1]{\texttt{#1}}
\providecommand{\urlprefix}{URL }
\providecommand{\doi}[1]{https://doi.org/#1}

\bibitem{bao2023one}
Bao, F., Nie, S., Xue, K., Li, C., Pu, S., Wang, Y., Yue, G., Cao, Y., Su, H., Zhu, J.: One transformer fits all distributions in multi-modal diffusion at scale. arXiv preprint arXiv:2303.06555  (2023)

\bibitem{campello2021multi}
Campello, V.M., Gkontra, P., Izquierdo, C., Martin-Isla, C., Sojoudi, A., Full, P.M., Maier-Hein, K., Zhang, Y., He, Z., Ma, J., et~al.: Multi-centre, multi-vendor and multi-disease cardiac segmentation: the m\&ms challenge. IEEE Transactions on Medical Imaging  \textbf{40}(12),  3543--3554 (2021)

\bibitem{chen2023generative}
Chen, B., Thandiackal, K., Pati, P., Goksel, O.: Generative appearance replay for continual unsupervised domain adaptation. arXiv preprint arXiv:2301.01211  (2023)

\bibitem{douillard2021plop}
Douillard, A., Chen, Y., Dapogny, A., Cord, M.: Plop: Learning without forgetting for continual semantic segmentation. In: Proceedings of the IEEE/CVF conference on computer vision and pattern recognition. pp. 4040--4050 (2021)

\bibitem{ho2020denoising}
Ho, J., Jain, A., Abbeel, P.: Denoising diffusion probabilistic models. Advances in neural information processing systems  \textbf{33},  6840--6851 (2020)

\bibitem{ho2022classifier}
Ho, J., Salimans, T.: Classifier-free diffusion guidance. arXiv preprint arXiv:2207.12598  (2022)

\bibitem{huang2023stu}
Huang, Z., Wang, H., Deng, Z., Ye, J., Su, Y., Sun, H., He, J., Gu, Y., Gu, L., Zhang, S., et~al.: Stu-net: Scalable and transferable medical image segmentation models empowered by large-scale supervised pre-training. arXiv preprint arXiv:2304.06716  (2023)

\bibitem{kirkpatrick2017overcoming}
Kirkpatrick, J., Pascanu, R., Rabinowitz, N., Veness, J., Desjardins, G., Rusu, A.A., Milan, K., Quan, J., Ramalho, T., Grabska-Barwinska, A., et~al.: Overcoming catastrophic forgetting in neural networks. Proceedings of the national academy of sciences  \textbf{114}(13),  3521--3526 (2017)

\bibitem{li2022domain}
Li, K., Yu, L., Heng, P.A.: Domain-incremental cardiac image segmentation with style-oriented replay and domain-sensitive feature whitening. IEEE Transactions on Medical Imaging  \textbf{42}(3),  570--581 (2022)

\bibitem{li2017learning}
Li, Z., Hoiem, D.: Learning without forgetting. IEEE transactions on pattern analysis and machine intelligence  \textbf{40}(12),  2935--2947 (2017)

\bibitem{liu2023clip}
Liu, J., Zhang, Y., Chen, J.N., Xiao, J., Lu, Y., A~Landman, B., Yuan, Y., Yuille, A., Tang, Y., Zhou, Z.: Clip-driven universal model for organ segmentation and tumor detection. In: Proceedings of the IEEE/CVF International Conference on Computer Vision. pp. 21152--21164 (2023)

\bibitem{liu2022learning}
Liu, P., Wang, X., Fan, M., Pan, H., Yin, M., Zhu, X., Du, D., Zhao, X., Xiao, L., Ding, L., et~al.: Learning incrementally to segment multiple organs in a ct image. In: International Conference on Medical Image Computing and Computer-Assisted Intervention. pp. 714--724. Springer (2022)

\bibitem{liu2020ms}
Liu, Q., Dou, Q., Yu, L., Heng, P.A.: Ms-net: multi-site network for improving prostate segmentation with heterogeneous mri data. IEEE transactions on medical imaging  \textbf{39}(9),  2713--2724 (2020)

\bibitem{liu2023incremental}
Liu, X., Shih, H.A., Xing, F., Santarnecchi, E., El~Fakhri, G., Woo, J.: Incremental learning for heterogeneous structure segmentation in brain tumor mri. In: International Conference on Medical Image Computing and Computer-Assisted Intervention. pp. 46--56. Springer (2023)

\bibitem{ma2024segment}
Ma, J., He, Y., Li, F., Han, L., You, C., Wang, B.: Segment anything in medical images. Nature Communications  \textbf{15}(1), ~654 (2024)

\bibitem{muller2022diffusion}
M{\"u}ller-Franzes, G., Niehues, J.M., Khader, F., Arasteh, S.T., Haarburger, C., Kuhl, C., Wang, T., Han, T., Nebelung, S., Kather, J.N., et~al.: Diffusion probabilistic models beat gans on medical images. arXiv preprint arXiv:2212.07501  (2022)

\bibitem{price2019privacy}
Price, W.N., Cohen, I.G.: Privacy in the age of medical big data. Nature medicine  \textbf{25}(1),  37--43 (2019)

\bibitem{radford2021learning}
Radford, A., Kim, J.W., Hallacy, C., Ramesh, A., Goh, G., Agarwal, S., Sastry, G., Askell, A., Mishkin, P., Clark, J., et~al.: Learning transferable visual models from natural language supervision. In: International conference on machine learning. pp. 8748--8763. PMLR (2021)

\bibitem{rombach2022high}
Rombach, R., Blattmann, A., Lorenz, D., Esser, P., Ommer, B.: High-resolution image synthesis with latent diffusion models. In: Proceedings of the IEEE/CVF conference on computer vision and pattern recognition. pp. 10684--10695 (2022)

\bibitem{song2020denoising}
Song, J., Meng, C., Ermon, S.: Denoising diffusion implicit models. arXiv preprint arXiv:2010.02502  (2020)

\bibitem{vaswani2017attention}
Vaswani, A., Shazeer, N., Parmar, N., Uszkoreit, J., Jones, L., Gomez, A.N., Kaiser, {\L}., Polosukhin, I.: Attention is all you need. Advances in neural information processing systems  \textbf{30} (2017)

\bibitem{wang2020dofe}
Wang, S., Yu, L., Li, K., Yang, X., Fu, C.W., Heng, P.A.: Dofe: Domain-oriented feature embedding for generalizable fundus image segmentation on unseen datasets. IEEE Transactions on Medical Imaging  \textbf{39}(12),  4237--4248 (2020)

\bibitem{wu2023continual}
Wu, H., Wang, Z., Zhao, Z., Chen, C., Qin, J.: Continual nuclei segmentation via prototype-wise relation distillation and contrastive learning. IEEE Transactions on Medical Imaging  (2023)

\bibitem{zhang2023s}
Zhang, J., Gu, R., Xue, P., Liu, M., Zheng, H., Zheng, Y., Ma, L., Wang, G., Gu, L.: S3r: Shape and semantics-based selective regularization for explainable continual segmentation across multiple sites. IEEE Transactions on Medical Imaging  (2023)

\bibitem{zhang2023jointnet}
Zhang, J., Li, S., Lu, Y., Fang, T., McKinnon, D., Tsin, Y., Quan, L., Yao, Y.: Jointnet: Extending text-to-image diffusion for dense distribution modeling. arXiv preprint arXiv:2310.06347  (2023)

\bibitem{zhang2022learning}
Zhang, J., Xue, P., Gu, R., Gu, Y., Liu, M., Pan, Y., Cui, Z., Huang, J., Ma, L., Shen, D.: Learning towards synchronous network memorizability and generalizability for continual segmentation across multiple sites. In: International Conference on Medical Image Computing and Computer-Assisted Intervention. pp. 380--390. Springer (2022)

\bibitem{zhao2023inherit}
Zhao, D., Yuan, B., Shi, Z.: Inherit with distillation and evolve with contrast: Exploring class incremental semantic segmentation without exemplar memory. IEEE Transactions on Pattern Analysis and Machine Intelligence  (2023)

\end{thebibliography}

\end{document}